\documentclass[aps,prd,a4paper,preprintnumbers,superscriptaddress,twocolumn,floatfix,showpacs,nofootinbib,usenatbib]{revtex4-1}
\usepackage{float,graphicx,multirow,bm,color,amsmath,amssymb,hyperref}
\hypersetup{
    colorlinks=true,
    linkcolor=black,
    citecolor=black,
}

\newcommand{\m}{{\rm M}}

\newcommand{\fR}{f_{R}}
\newcommand{\dfR}{{\delta}f_{R}}

\newcommand{\rd}{{\rm d}}
\newcommand{\dr}{\delta{R}}
\newcommand{\remove}[1]{}

\def\be{\begin{equation}}
\def\ee{\end{equation}}
\def\ba{\begin{eqnarray}}
\def\ea{\end{eqnarray}}

\begin{document}

\title{{\tt ECOSMOG}: An Efficient Code for Simulating Modified Gravity}
\author{Baojiu~Li}
\email[Email address: ]{b.li@damtp.cam.ac.uk}
\affiliation{ICC, Physics Department, University of Durham, South Road, Durham DH1 3LE, UK}
\affiliation{DAMTP, Centre for Mathematical Sciences, University of Cambridge, Wilberforce Road, Cambridge CB3 0WA, UK}
\affiliation{Kavli Institute for Cosmology Cambridge, Madingley Road, Cambridge CB3 0HA, UK}
\affiliation{Institute of Astronomy, Madingley Road, Cambridge CB3 0HA, UK}

\author{Gong-Bo~Zhao}
\email[Email address: ]{gong-bo.zhao@port.ac.uk}
\affiliation{Institute of Cosmology and Gravitation, University of Portsmouth,
Portsmouth, PO1 3FX, UK}

\author{Romain~Teyssier}
\email[Email address: ]{teyssier@physik.uzh.ch}
\affiliation{Universit\"at Z\"urich, Institute f\"ur Theoretische Physik, Winterthurerstrasse 190,
CH-8057 Z\"urich, Switzerland}
\affiliation{CEA Saclay, DSM/IRFU/SAP, B\^atiment 709, F-91191 Gif-sur-Yvette, Cedex,
France}

\author{Kazuya~Koyama}
\email[Email address: ]{kazuya.koyama@port.ac.uk}
\affiliation{Institute of Cosmology and Gravitation, University of Portsmouth,
Portsmouth, PO1 3FX, UK}

\date{\today}

\begin{abstract}
We introduce a new code, {\tt ECOSMOG}, to run $N$-body simulations for a wide class of modified gravity and dynamical dark energy theories. These theories generally have one or more new dynamical degrees of freedom, the dynamics of which are governed by their (usually rather nonlinear) equations of motion. Solving these non-linear equations has been a great challenge in cosmology. Our code is based on the {\tt RAMSES} code, which solves the Poisson equation on adaptively refined meshes to gain high resolutions in the high-density regions. We have added a solver for the extra degree(s) of freedom and performed numerous tests for the $f(R)$ gravity model as an example to show its reliability. We find that much higher efficiency could be achieved compared with other existing mesh/grid-based codes thanks to two new features of the present code: (1) the efficient parallelisation and (2) the usage of the multigrid relaxation to solve the extra equation(s) on {\it both} the regular domain grid {\it and} refinements, giving much faster convergence even under much more stringent convergence criteria. This code is designed for performing high-accuracy, high-resolution and large-volume cosmological simulations for modified gravity and general dark energy theories, which can be utilised to test gravity and the dark energy hypothesis using the upcoming and future deep and high-resolution galaxy surveys.

\end{abstract}


\maketitle

\section{Introduction}

The accelerating expansion of our Universe is one of the most challenging questions in modern physics \cite{cst2006}. After more than a decade of attempts in constructing consistent and natural theories for it, we are still nowhere near a definite conclusion. Broadly speaking, the theoretical models developed so far roughly fall into two categories: those involving some exotic matter species, the so-called dark energy, which usually have nontrivial dynamics, and those involving modifications to the Einstein gravity on certain (usually large) scales. Examples of the dynamical dark energy include the quintessence \cite{zws1999}, k-essence {ams2000}, coupled quintessence \cite{a2000}, chameleon field \cite{kw2004,ms2007}, symmetron field \cite{hk2010} {\it etc.}, and some examples of modified gravity models are the $f(R)$ gravity \cite{cddett2005}, scalar-tensor theory \cite{pb1999} and DGP model \cite{dgp2000} (for some recent review see, {\it e.g.}, \cite{k2008,dm2008,sf2008}).

From a practical point of view ({\it i.e.}, regardless of the naturalness or fine-tuning considerations), there are two important issues faced by any theoretical model: consistency and degeneracy.

In order to ensure the consistency of the mode, a given model should not violate any existing observational constraints. The new degrees of freedom, which are supposed to drive the accelerating expansion only on very large scales, quite often produce unwanted side effects. Consider the coupled quintessence as an example; if it couples to normal matter species (baryons {\it etc.}), it can mediate a fifth force that is strongly constrained by experiments. This problem can certainly be avoided by assuming that the quintessence field only couples to dark matter which we cannot directly observe, but more interesting theoretical mechanisms to avoid this problem have been designed in recent years. The leading example is the chameleon mechanism \cite{kw2004,ms2007,lb2007}. By this mechanism, the fifth force is suppressed to the undetectable level in regions with high matter density, while in low-density environments such as galaxies, galaxy clusters and cosmological voids, the fifth force is unsuppressed and can be as strong as gravity. The rapid changes of the behaviour of the fifth force across different regions naturally imply that the equation(s) of motion governing the dynamics of the new degree(s) of freedom should be very nonlinear. This adds the complexity to the model but we cannot avoid this to ensure the consistency of the model.

Different theoretical models can behave very similarly to each other and to the standard $\Lambda$CDM paradigm, which makes it difficult to distinguish amongst them using observational data. The (apparent) existence of the degeneracy usually indicates the incompleteness of our theoretical understanding of models. For instance, it is very easy to design a scalar field model which produces almost identical cosmic expansion history as $\Lambda$CDM, but this is merely because in the background cosmology, any detailed structure of the Universe is disregarded; when we look at the linear perturbation evolutions, the apparent degeneracy is often broken. However, there are situations where the degeneracy cannot be broken even with linear perturbation analysis (such as the chameleon theory and $f(R)$ gravity in which the length scales on which linear perturbation analyses apply are bigger than the range of, and so not affected by, the fifth force). In such situations, a full study of the nonlinear structure formation and evolution needs to be performed to break the degeneracy.

Therefore, it is essential to study non-linea structure formation to ensure the consistency of the model and break the degeneracies between various models. Non-linear structure formation is too complex to understand analytically and therefore requires numerical simulations. Numerical simulations of the cosmic structure evolution on small (such as galactic or cluster) scales can not only open a new arena of using consistency with observations to constrain models, but also hopefully break the degeneracy between different models (as our cosmological simulations show below). This is particularly exciting considering that high-quality observational data will keep coming in the next decades to improve our theoretical understanding.

Numerical simulations are done by numerical codes. In most cases the extra dynamical degree(s) of freedom is more accurately treated as a field and its value is required on a number of space points. For this purpose we need the numerical code that can solve the field value on meshes covering (parts of) the simulation box. There are two such codes known to us to date: one is that of Oyaizu \cite{o2008}, which has been applied to $f(R)$ gravity \cite{olh2008,sloh2009} and the DGP model \cite{f2009}; the other is a modified {\tt MLAPM} code \cite{mlapm} written by one of the authors \cite{lz2009} and which has been applied to chameleon theories \cite{lz2010}, $f(R)$ gravity \cite{zlk2011}, coupled scalar field theories \cite{lb2011,lmb2011a}, scalar-tensor theory \cite{lmb2011b}, dilaton model \cite{bbdls2011}, symmetron model \cite{dlmw2011}. Both have shortcomings: the Oyaizu's code does not support adaptive refinements (thus easier to implement) and high-resolution simulations are not practical; while the {\tt MLAPM} code does support adaptive refinements, it is not parallelised and not practical for simulations with very big volumes and high mass resolutions. Furthermore, the equations on {\tt MLAPM} refinements are solved on a one-level grid, which is rather inefficient.

The aim of this paper is to introduce a new code {\tt ECOSMOG}, which overcomes the shortcomings of the previous codes. The new code is based on {\tt RAMSES} \cite{ramses}. It is efficiently parallelised, supports adaptive refinements and solves the equations using multigrid method on the refinements (for a more detailed comparison of the three codes the readers can refer to the table I). As a working example, we use the new code to run a number of test simulations for the $f(R)$ gravity, which is one of the most challenging models to simulate because of the high nonlinearity of its equations. As will be shown below, the code works very well for the $f(R)$ model, and we certainly expect it to be straightforward to implement equations in other models to our code.

The organisation of this paper is as follows. In \S~\ref{sect:fr_gravity} we briefly introduce the $f(R)$ gravity model. In \S~\ref{sect:nbody_eqns} we introduce the supercomoving code unit using a different form and list the $N$-body Poisson and $f(R)$ equations to simplify the numerics. \S~\ref{sect:discrete_eqns} then makes discrete versions of these equations to be implemented in our code. \S~\ref{sect:nbody_algorithm}, we show details on how the numerical implementation is performed and discuss several important issues, which is the core section of this paper. Next, a long section, \S~\ref{sect:tests}, contains the results of eleven tests of the code. These tests check the correctness, efficiency and consistency of different aspects of the code and give us confidence about its reliability. Finally we compare the present code with other mesh-based codes, summarise and conclude in \S~\ref{sect:summary}.

This is a paper to explain the code and physical interpretations of the results will be presented in future publications.

\section{A Test Case: The $f(R)$ Gravity}

\label{sect:fr_gravity}

One can alter the Einstein gravity in such a way that it gives rise the cosmic acceleration without introducing dark energy. One example along this line is the so-called $f(R)$ gravity, in which the Ricci scalar $R$ in the Einstein-Hilbert action is generalised to a function of $R$ (see {\it e.g.}, \cite{sf2008} for a review and references therein). In $f(R)$ gravity, the structure formation is governed by the following two equations,

\begin{eqnarray}
\label{eq:Poisson} \nabla^2 \Phi&=&\frac{16\pi{G}}{3}a^2\delta\rho_{\m}+\frac{a^2}{6}\dr(\fR),\\
\label{eq:sf} \nabla^2\dfR &=&-\frac{a^2}{3}[\dr(\fR)+8\pi{G}\delta\rho_{\m}],
\end{eqnarray}
where $\Phi$ denotes the gravitational potential, $\fR\equiv\frac{\rd f(R)}{\rd R} $ is the extra scalar degree of freedom, dubbed \emph{scalaron}, $\dfR=\fR(R)-\fR(\bar{R}),\dr=R-\bar{R},\delta\rho_{\m}=\rho_{\m}-\bar{\rho_{\m}}$, and the quantities with overbar take the background values. The symbol $\nabla$ is the three dimensional gradient operator, and $a$ is the scale factor. These two coupled Poisson-like equations are much difficult to solve than the single Poisson equation in General Relativity (GR), $\nabla^2 \Phi=4\pi{G}a^2\delta\rho_{\m}$, which is linear. From Eqs (\ref{eq:Poisson}) and (\ref{eq:sf}), we can see that gravity in $f(R)$ can be enhanced depending on the local environment -- in underdense regions, the $\dr(\fR)$ term in Eqs (\ref{eq:Poisson}) vanishes thus the two equations decouple, making gravity simply enhanced by a factor of $4/3$. However in the dense region, $\dfR$ in Eq (\ref{eq:sf}) is negligible, yielding $\dr(\fR)=-8\pi{G}\delta\rho_{\m}$, which means that GR is locally restored. This is the chameleon mechanism applied in the $f(R)$ gravity models, which is important for the cosmological viability of the latter. The presence of the chameleon effect indicates that Eqs (\ref{eq:Poisson}) and (\ref{eq:sf}) are highly nonlinear, making it challenging to numerically solve them in the simulation process.

An $f(R)$ gravity model is fully specified by the functional form of $f(R)$, and here we shall adopt the model proposed by Hu \& Sawicki \cite{hs2007}, which takes the form
\begin{eqnarray}
f(R) &=& -m^2\frac{c_1(-R/m^2)^n}{c_2(-R/m^2)^n+1},
\end{eqnarray}
where $n, c_1, c_2$ are model parameters, and 
\begin{eqnarray}
m^2 &\equiv& \Omega_mH_0^2,
\end{eqnarray}
with $\Omega_m$ being the present fractional matter density and $H_0$ the current Hubble expansion rate.

It can then be shown that
\begin{eqnarray}
f_R &=& -\frac{c_1}{c_2^2}\frac{n(-R/m^2)^{n+1}}{[(-R/m^2)^n+1]^2}.%\nonumber\\
%&\approx& -n\frac{c_1}{c_2^2}\left[\frac{m^2}{-R}\right]^{n+1}
\end{eqnarray}
Because 
\begin{eqnarray}
-\bar{R}\approx8\pi G\bar{\rho}_m-2\bar{f}(R)=3m^2\left[a^{-3}+\frac{2}{3}\frac{c_1}{c_2}\right],
\end{eqnarray}
where overbars are used for the background quantities, to match a $\Lambda$CDM background expansion we have $c_1/c_2=6\Omega_\Lambda/\Omega_m$. With $\Omega_m=0.24$ and $\Omega_\Lambda=0.76$, we find $-\bar{R}\approx41m^2\gg m^2$, which means that $f_R$ can be approximated as
\begin{eqnarray}
f_R &\approx& -n\frac{c_1}{c_2^2}\left[\frac{m^2}{-R}\right]^{n+1}.
\end{eqnarray}
Because $f_R$ is the actual quantity that enters the $N$-body equations (see below), we find that only the two parameters $n$ and $\xi\equiv c_1/c_2^2$, are needed in our simulations. Another (independent) parameter $f_{R0}$ which is the present background value of $f_R$, can be obtained from $\xi$ and is often used to give people some idea about the size of $f_R$.

\section{The $N$-body Equations}

In this section we shall introduce the convention of our code units, and list the $N$-body Poisson and $f(R)$ equations, the latter being written in the so-called quasi-static limit so that terms involving time derivatives will be dropped. The $N$-body equations can all be found elsewhere, though perhaps of slightly different forms; consequently, this section shall be kept short and only serves for completeness.

\label{sect:nbody_eqns}

\subsection{The Code Units}

The code unit used in the {\tt RAMSES} code and its modification developed here is based on (but not exactly) the supercomoving coordinates of \cite{ms1998}. It can be summarised as follows (where tilded quantities are expressed in the code unit):
\begin{eqnarray}
\tilde{x}\ =\ \frac{x}{aB},\ \ \ \tilde{\rho}\ =\ \frac{\rho a^3}{\rho_c\Omega_m},\ \ \ \tilde{v}\ =\ \frac{av}{BH_0},\nonumber\\
\tilde{\phi}\ =\ \frac{a^2\phi}{(BH_0)^2},\ \ \ d\tilde{t}\ =\ H_0\frac{dt}{a^2},\ \ \ \tilde{c}\ =\ \frac{c}{BH_0}.\nonumber
\end{eqnarray}
In the above $x$ is the comoving coordinate, $\rho_c$ is the critical density today, $\Omega_m$ the fractional energy density for matter today, $v$ the particle velocity, $\phi$ the gravitational potential and $c$ the speed of light. In addition, $B$ is the size of the simulation box in unit of $h^{-1}$Mpc and $H_0$ the Hubble expansion rate today in unit of $100h$~km/s/Mpc. Note that with these conventions the average matter density is $\tilde{\bar{\rho}}=1$. All the newly-defined quantities are dimensionless.

In the $f(R)$ gravity equation we also have a new degree of freedom $f_R$ and in the code unit we will use $\tilde{f}_R\equiv a^2f_R$ instead. As we shall see below, this is to make sure that $\tilde{f}_R$ has an equivalent status as $\tilde{\phi}$: the latter is the Newtonian potential and determines the total (modified) gravitational force, while the former is related to the potential of the extra force and determines the modification to the standard Einsteinian gravity.
\subsection{The Modified Poisson Equation}

From above we see that the modified Poisson equation is
\begin{eqnarray}
\vec{\nabla}^2\phi &=& \frac{16\pi G}{3}\delta\rho - \frac{1}{6}\delta R
\end{eqnarray}
where $\delta R=R-\bar{R}$ and
\begin{eqnarray}
R &=& m^2\left(-\frac{n\xi}{f_R}\right)^{\frac{1}{n+1}},\\
\bar{R} &=& 3m^2\left(a^{-3}+4\frac{\Omega_\Lambda}{\Omega_m}\right).
\end{eqnarray}

Using the code units defined above, the equation becomes
\begin{eqnarray}
\tilde{\nabla}^2\tilde{\phi} &=& 2\Omega_ma\left(\tilde{\rho}-1\right)\nonumber\\
&&- \frac{1}{6}\Omega_ma^4\left[\left(-\frac{na^2\xi}{\tilde{f}_R}\right)^{\frac{1}{n+1}} - 3\left(a^{-3}+4\frac{\Omega_\Lambda}{\Omega_m}\right)\right].
\end{eqnarray}

\subsection{The $f(R)$ Equation}

The equation of motion for $f_R$,
\begin{eqnarray}
\vec{\nabla}^2f_R &=& \frac{1}{3c^2}\left[\delta R-8\pi G\delta\rho\right],
\end{eqnarray}
becomes in the code units,
\begin{eqnarray}
&&\tilde{\nabla}^2\tilde{f}_R\nonumber\\
&=& -\frac{1}{3\tilde{c}^2}\Omega_ma\left(\tilde{\rho}-1\right)\nonumber\\
&&+\frac{1}{3\tilde{c}^2}\Omega_ma^4\left[\left(-\frac{na^2\xi}{\tilde{f}_R}\right)^{\frac{1}{n+1}} - 3\left(a^{-3}+4\frac{\Omega_\Lambda}{\Omega_m}\right)\right].
\end{eqnarray}
Comparing this equation with that for the modified Poisson equation, we find that $\tilde{\phi}$ and $\tilde{c}^2\tilde{f}_R$ has the same code unit, and that is why we have used $\tilde{f}_R$ instead of $f_R$.

As in \cite{zlk2011}, we will not solve $\tilde{f}_R$ directly as it may change too quickly from one space point to another and can cause divergence problems in the numerical solutions. Instead, we will solve a new variable $\tilde{u}$, defined through $\tilde{f}_R\equiv-e^{\tilde{u}}$. $\tilde{u}$ will be of order unity across the space, and so has better convergence properties.

\section{The Discretised Equations}

\label{sect:discrete_eqns}

From here on we will only use variables expressed using the code unit, and the tildes will be dropped for simplicity.

Clearly, to put the above equations into the $N$-body code one must discretise them. For the Poisson equation, it is linear as along as the source term (the right-hand side) is provided, so the discretisation is fairly straightforward:
\begin{widetext}
\begin{eqnarray}\label{eq:discrete_poisson}
&&\frac{1}{h^2}\left[\phi_{i+1,j,k}+\phi_{i-1,j,k}+\phi_{i,j+1,k}+\phi_{i,j-1,k}+\phi_{i,j,k+1}+\phi_{i,j,k-1}-6\phi_{i,j,k}\right]\nonumber\\
&=& 2\Omega_ma\left(\delta_{i,j,k}-1\right)- \frac{1}{6}\Omega_ma^4\left[(na^2\xi)^{\frac{1}{n+1}}\exp\left(-\frac{u_{i,j,k}}{n+1}\right) - 3\left(a^{-3}+4\frac{\Omega_\Lambda}{\Omega_m}\right)\right],
\end{eqnarray}
\end{widetext}
where for example $\phi_{i,j,k}$ is the value of $\phi$ in the grid cell with index $(i,j,k)$. The discrete version of the nonlinear $f(R)$ equation of motion, in contrast, is more tedious \cite{zlk2011}:
\begin{widetext}
\begin{eqnarray}\label{eq:discrete_fR}
&&\frac{1}{h^2}\left[b_{i+\frac{1}{2},j,k}u_{i+1,j,k}-u_{i,j,k}\left(b_{i+\frac{1}{2},j,k}+b_{i-\frac{1}{2},j,k}\right)+b_{i-\frac{1}{2},j,k}u_{i-1,j,k}\right]\nonumber\\
&&+\frac{1}{h^2}\left[b_{i,j+\frac{1}{2},k}u_{i,j+1,k}-u_{i,j,k}\left(b_{i,j+\frac{1}{2},k}+b_{i,j-\frac{1}{2},k}\right)+b_{i,j-\frac{1}{2},k}u_{i,j-1,k}\right]\nonumber\\
&&+\frac{1}{h^2}\left[b_{i,j,k+\frac{1}{2}}u_{i,j,k+1}-u_{i,j,k}\left(b_{i,j,k+\frac{1}{2}}+b_{i,j,k-\frac{1}{2}}\right)+b_{i,j,k-\frac{1}{2}}u_{i,j,k-1}\right]\nonumber\\
&&+\frac{1}{3\tilde{c}^2}\Omega_ma^4\left(na^2\xi\right)^{\frac{1}{n+1}}\exp\left(-\frac{u_{i,j,k}}{n+1}\right) - \frac{1}{\tilde{c}^2}\Omega_ma\left(\delta_{i,j,k}-1\right)-\frac{1}{\tilde{c}^2}\Omega_ma^4\left(a^{-3}+4\frac{\Omega_\Lambda}{\Omega_m}\right)\ =\ 0.
\end{eqnarray}
\end{widetext}
Notice that in the above equations we have used $\delta_{i,j,k}\equiv\tilde{\rho}_{i,j,k}$ to avoid confusion in notations and to make it clear that we are using the density {\it contrast}. $b\equiv e^u$ and
\begin{eqnarray}
b_{i+\frac{1}{2},j,k} &\equiv& \frac{1}{2}\left(b_{i+1,j,k}+b_{i,j,k}\right),\nonumber\\
b_{i-\frac{1}{2},j,k} &\equiv& \frac{1}{2}\left(b_{i,j,k}+b_{i-1,j,k}\right),~\cdots\nonumber
\end{eqnarray}

\section{The $N$-body Algorithms}

\label{sect:nbody_algorithm}

The detailed implementation of the $N$-body solvers in the {\tt RAMSES} code is indeed quite different from that in the {\tt MLAPM} code \cite{lz2009,lb2011}. As a result, certain corrections need to be made to the above discrete equations before implementing them.

Both codes adaptively refine the grid in high-density regions and solve the Poisson and $f(R)$ equations on the refinement to get better spatial resolutions. The refining is performed on a grid-by-grid basis so that the final refinements typically have irregular shapes roughly matching the iso-density surfaces. Particles in regions underlying the refinements are linked to the latter, where their positions and momenta are updated.

An important difference between the two codes is how the equations are solved on the refinements. {\tt MLAPM} solves the equations on single level of the considered refinement only, performing Gauss-Seidel relaxations till the convergence of the solution is obtained. In {\tt RAMSES}, however, for each adaptive mesh refinement (AMR) level there is a series of separate, coarser multigrid levels on which the Gauss-Seidel relaxation is used \remove{in conjunction with the multigrid technique} to accelerate the convergence.

The treatments of boundary conditions on refinements are also different. In {\tt MLAPM}, the outermost cells of a refinement are taken as its physical boundary, thus the field values wherein are set using interpolation from the coarser-levels solutions. In {\tt RAMSES}, the use of multilevel requires the boundary conditions to be set consistently on the different levels, and this is achieved using an elegant {\it mask} scheme: on the AMR level, which is the finest level in the multigrid series, the active cells, inside which the field values need to be updated during the relaxation process, are given a mask value of $+1$ while the other cells are assigned a mask value $-1$. The boundary of the computational domain is defined to be where the linearly-interpolated mask value vanishes, or equally the boundary of the active AMR cells. The value of the field on this physical boundary is computed at the beginning of the relaxation process and kept unchanged after that until a converged solution in the active cells are obtained. The mask values in the coarser multigrid cells are restricted from their finer-level values, while the physical boundary on this level is still defined as where the mask hits zero, which ensures the boundaries on different levels are consistent (for more details and tests in the classical GR case see \cite{gt2011}).

Moreover, the treatments of the boundary conditions for linear and nonlinear elliptical PDEs are also different in {\tt RAMSES}, as we shall see below.

\subsection{The Modified Poisson Equation}

\subsubsection{The BC-modified Equation}

Unlike the {\tt MLAPM} code, where the refinement boundary consists of the outermost cells on that refinement and so the boundary cells are always there, in {\tt RAMSES} we might well face the situation that an outermost active cell has no neighbouring cells in some directions on the save level (possibly because a neighbouring coarse cell has not been refined), while we do need the field values in those cells to interpolate and compute the boundary conditions. Such cells are called ghost cells: they do not exist in the code data structure but we do need them.

For the linear elliptical PDEs, such as the (modified) Poisson equation, there is a simple way to avoid storing information about the ghost cells. Consider the discrete Poisson equation and now suppose that the $(i-1,j,k)$-cell is a ghost, which has mask value $m_{i-1,j,k}$\footnote{If this is a ghost cell then its mask value will be $-1$ but here we try to be general in our description.}. We realise that the physical boundary, namely where the mask value crosses zero, will be a distance
\begin{eqnarray}
\frac{-m_{i-1,j,k}}{m_{i,j,k}-m_{i-1,j,k}}h\nonumber
\end{eqnarray}
from the $(i-1,j,k)$ (ghost!) cell and
\begin{eqnarray}
\frac{m_{i,j,k}}{m_{i,j,k}-m_{i-1,j,k}}h\nonumber
\end{eqnarray}
from the $(i,j,k)$-cell, where $h$ is the cell length, or equally the distance between the centres of these two cells. Using linear interpolation, the boundary value of $\phi$ will then be
\begin{eqnarray}
\phi_b &=& \frac{m_{i,j,k}}{m_{i,j,k}-m_{i-1,j,k}}\phi_{i-1,j,k}-\frac{m_{i-1,j,k}}{m_{i,j,k}-m_{i-1,j,k}}\phi_{i,j,k},\nonumber
\end{eqnarray}
which gives
\begin{eqnarray}\label{eq:phi_b}
\phi_{i-1,j,k} &=& \phi_b - \frac{m_{i-1,j,k}}{m_{i,j,k}}\phi_b+\frac{m_{i-1,j,k}}{m_{i,j,k}}\phi_{i,j,k}.
\end{eqnarray}
Note that though the $(i-1,j,k)$-cell is a ghost cell, the value of $\phi_{i-1,j,k}$ can be computed, at the beginning of the relaxation iterations, from its father (coarser) cell which does exist. Then $\phi_b$ is computed and fixed (because it is the boundary value!). During the subsequent relaxation iterations, the value of $\phi_{i,j,k}$ changes and so does that of $\phi_{i-1,j,k}$ but not $\phi_b$. As we don't have the $(i-1,j,k)$-cell, there is nowhere to store the updated values of $\phi_{i-1,j,k}$, and so we choose to not use it at all, replacing $\phi_{i-1,j,k}$ in the discrete equation by $\phi_b$ and $\phi_{i,j,k}$ using Eq.~(\ref{eq:phi_b}). We then get a boundary-condition (BC)-modified equation
\begin{widetext}
\begin{eqnarray}\label{eq:bc_modified_poisson}
&&\frac{1}{h^2}\left[\phi_{i+1,j,k}+\phi_{i,j+1,k}+\phi_{i,j-1,k}+\phi_{i,j,k+1}+\phi_{i,j,k-1}-\left(6-\frac{m_{i-1,j,k}}{m_{i,j,k}}\right)\phi_{i,j,k}\right]\nonumber\\
&=& 2\Omega_ma\left(\delta_{i,j,k}-1\right)- \frac{1}{6}\Omega_ma^4\left[(na^2\xi)^{\frac{1}{n+1}}\exp\left(-\frac{u_{i,j,k}}{n+1}\right) - 3\left(a^{-3}+4\frac{\Omega_\Lambda}{\Omega_m}\right)\right]
-\frac{1}{h^2}\left(1-\frac{m_{i-1,j,k}}{m_{i,j,k}}\right)\phi_b.
\end{eqnarray}
\end{widetext}
Note that because $\phi_b$ is fixed, we could simply move it to the right-hand side of the equation. The BC-modified right-hand side is then only needed to be computed once, before entering the relaxation iterations. There is no need to store either $\phi_b$ or $\phi_{i-1,j,k}$. Note that the left-hand side is also modified and one must be careful here.

\subsubsection{The Multigrid Implementation}

Now consider the multigrid algorithm. For simplicity let us rewrite the above BC-modified equation as
\begin{eqnarray}
L^h\phi^h &=& f^h
\end{eqnarray}
in which the superscript means we are on the level with cell size $h$. Note that $L^h$ is a linear operator, which is important here.

Suppose after a number of Gauss-Seidel iterations (pre-smoothing), we get an approximate solution $\hat{\phi}^h$, then
\begin{eqnarray}
L^h\hat{\phi}^h\ =\ \hat{f}^h\ \neq\ f^h
\end{eqnarray}
in the active cells, and the difference
\begin{eqnarray}
d^h &\equiv& \hat{f}^h-f^h
\end{eqnarray}
is called the residual. We could define $\delta\phi^h\equiv\phi^h-\hat{\phi}^h$ and rewrite the equation as
\begin{eqnarray}
L^h\delta\phi^h &=& -d^h
\end{eqnarray}
and coarsify this equation as
\begin{eqnarray}
L^H\delta\phi^H &=& -\mathcal{R}d^h
\end{eqnarray}
and solve it on the coarser level to accelerate the convergence. Here, the superscript $^H$ indicates that we are on the coarser level on which the cell size is $H=2h$, and $\mathcal{R}$ is the restriction operator.

To solve this coarsified equation, we need the boundary conditions for $\delta\phi^H$ on the coarser level and, as in Eq.~(\ref{eq:bc_modified_poisson}), the coarser-level equation will be modified to include correction terms involving $\delta\phi_B$, which is the boundary $\delta\phi$ on the coarser level. However, on the boundary $\delta\phi$ vanishes identically and it turns out that the coarser-level version of Eq.~(\ref{eq:bc_modified_poisson}) is simpler:
\begin{widetext}
\begin{eqnarray}\label{eq:bc_modified_poisson_coarse}
\frac{1}{H^2}\left[\phi^H_{i+1,j,k}+\phi^H_{i,j+1,k}+\phi^H_{i,j-1,k}+\phi^H_{i,j,k+1}+\phi^H_{i,j,k-1}-\left(6-\frac{m^H_{i-1,j,k}}{m^H_{i,j,k}}\right)\phi^H_{i,j,k}\right] &=& -\mathcal{R}d^h,
\end{eqnarray}
\end{widetext}
assuming that the $(i-1,j,k)$ {\it coarser-level} cell is a ghost cell. Note that the right-hand side of the coarsified equation is {\it not} BC modified since $\delta\phi_B=0$, but the left-hand side {\it is} modified by the lack of neighbouring cells.

Here we have only described the algorithm for two levels, but the generalisation to more levels is trivial.

Finally some comments should be made on the restriction operation $\mathcal{R}d^h$: if a given fine cell is masked\footnote{Here "masked" means having a non-positive mask value.} then it has no contribution to the coarser-level residual, because the field value in this cell is unmodified by the relaxation and so considered to be exact. At the same time, we shall not restrict residuals into coarse cells which are masked, because this is unnecessary.

\subsubsection{The Prolongation}

Once an approximate coarser-level solution to $\delta\phi^H$ is obtained, say $\hat{\delta\phi}^H$, one can correct the fine-level solution using
\begin{eqnarray}
\hat{\phi}^h &\leftarrow& \hat{\phi}^h + \mathcal{P}\hat{\delta\phi}^H,
\end{eqnarray}
where $\mathcal{P}$ is the prolongation operator, to find (expected) more accurate solutions on the fine level.

In the prolongation process, a fine cell receives contribution from its eight neighbouring coarser cells. Of these eight coarse cells:
\begin{enumerate}
\item one contains the fine cell and is given a weight of $27/64$,
\item three have one common face with the fine cell and are given a weight of $9/64$ each,
\item three have one common edge with the fine cell and are given a weight of $3/64$ each,
\item one has a common vertex with the fine cell and is given a weight of $1/64$.
\end{enumerate}
Of course, these weights sum to unity.

Note that if a given fine cell is masked, then it is outside the active computational domain and corrections are not necessary for it. If the coarse cell is not a valid multigrid cell (it could be a valid AMR cell, though), then its contribution to the fine-cell correction is zero.

\subsection{The $f(R)$ Equation}

Having recalled the multigrid algorithm for the linear (modified) Poisson equation, first presented in \cite{gt2011}, we can now have a look at the nonlinear $f(R)$ equation. The nonlinearity introduces certain complications compared to the linear case, which should be taken care of in the numerical implementations.

\subsubsection{The BC-Modified Equation}

As in the linear case, let us suppose that the $(i-1,j,k)$-cell on the finest level is a ghost cell. Then
\begin{eqnarray}\label{eq:u_i-1jk}
u_{i-1,j,k} &=& u_b - \frac{m_{i-1,j,k}}{m_{i,j,k}}u_b + \frac{m_{i-1,j,k}}{m_{i,j,k}}u_{i,j,k}.
\end{eqnarray}
with $u_b$ the boundary value of $u$. Note that this necesarilly means that the equality
\begin{eqnarray}
b_{i-1,j,k} &=& b_b - \frac{m_{i-1,j,k}}{m_{i,j,k}}b_b + \frac{m_{i-1,j,k}}{m_{i,j,k}}b_{i,j,k}
\end{eqnarray}
does {\it not} hold, because $b=e^u$ is nonlinear in $u$ -- this point is very important in order to make consistent BC-modified $f(R)$ equation. Instead, we shall use $b_{i-1,j,k}\equiv\exp(u_{i-1,j,k})$ with $u_{i-1,j,k}$ given by Eq.~(\ref{eq:u_i-1jk}).

As before, $u_{i-1,j,k}$ is computed and then used to obtain $u_b$ before entering the relaxation iterations. In the subsequent iterations, $u_{i,j,k}$ is updated and so is $u_{i-1,j,k}$, but not $u_b$.

Removing the $u_{i-1,j,k}$ and $b_{i-1,j,k}$ in Eq.~(\ref{eq:discrete_fR}) using the above two relations, we arrive at the BC modified $f(R)$ equation:
\begin{widetext}
\begin{eqnarray}\label{eq:bc_modified_fR_op}
&&\frac{\tilde{c}^2}{2h^2}b_{i,j,k}\left[u_{i+1,j,k}+u_{i,j+1,k}+u_{i,j-1,k}+u_{i,j,k+1}+u_{i,j,k-1}+\left(1-\frac{m_{i-1,j,k}}{m_{i,j,k}}\right)u_b-\left(6-\frac{m_{i-1,j,k}}{m_{i,j,k}}\right)u_{i,j,k}\right]\nonumber\\
&-&\frac{\tilde{c}^2}{2h^2}u_{i,j,k}\left[b_{i+1,j,k}+b_{i,j+1,k}+b_{i,j-1,k}+b_{i,j,k+1}+b_{i,j,k-1}+\left(1-\frac{m_{i-1,j,k}}{m_{i,j,k}}\right)e^{\left(1-\frac{m_{i-1,j,k}}{m_{i,j,k}}\right)u_b}e^{\frac{m_{i-1,j,k}}{m_{i,j,k}}u_{i,j,k}}\right]\nonumber\\
&+&\frac{\tilde{c}^2}{2h^2}\left[b_{i+1,j,k}u_{i+1,j,k}+b_{i,j+1,k}u_{i,j+1,k}+b_{i,j-1,k}u_{i,j-1,k}+b_{i,j,k+1}u_{i,j,k+1}+b_{i,j,k-1}u_{i,j,k-1}\right]\nonumber\\
&+&\frac{\tilde{c}^2}{2h^2}\left(1-\frac{m_{i-1,j,k}}{m_{i,j,k}}\right)u_be^{\left(1-\frac{m_{i-1,j,k}}{m_{i,j,k}}\right)u_b}e^{\frac{m_{i-1,j,k}}{m_{i,j,k}}u_{i,j,k}}-\Omega_ma(\tilde{\rho}_{i,j,k}-1)\nonumber\\
&+&\frac{1}{3}\Omega_ma^4\left(na^2\xi\right)^{\frac{1}{n+1}}\exp\left(-\frac{u_{i,j,k}}{n+1}\right)-\Omega_ma^4\left(a^{-3}+4\frac{\Omega_\Lambda}{\Omega_m}\right)\ =\ 0.
\end{eqnarray}
\end{widetext}
Now we could see clearly where the additional complexity appears. Remember that in the linear case we don't have to store $\phi_b$ because it only appears on the BC-modified right-hand side of the equation. Here, on the other hand, $u_b$ also appears in the term $u_bb_{i,j,k}$ such that its value is needed in each iteration during which $b_{i,j,k}$ is updated. This implies that, for each outermost active cell, we should store $u_b$ for its neighbouring ghost cell(s).

If we define the the left-hand side of Eq.~(\ref{eq:bc_modified_fR_op}) as $\mathcal{L}^h$ for simplicity, then it can be easily shown that
\begin{widetext}
\begin{eqnarray}\label{eq:bc_modified_fR_dop}
&&\frac{\partial\mathcal{L}^h(u_{i,j,k})}{\partial u_{i,j,k}}\nonumber\\
&=& \frac{\tilde{c}^2}{2h^2}b_{i,j,k}\left[u_{i+1,j,k}+u_{i,j+1,k}+u_{i,j-1,k}+u_{i,j,k+1}+u_{i,j,k-1}+\left(1-\frac{m_{i-1,j,k}}{m_{i,j,k}}\right)u_b-\left(6-\frac{m_{i-1,j,k}}{m_{i,j,k}}\right)u_{i,j,k}\right]\nonumber\\
&&-\frac{\tilde{c}^2}{2h^2}\left[b_{i+1,j,k}+b_{i,j+1,k}+b_{i,j-1,k}+b_{i,j,k+1}+b_{i,j,k-1}+\left(1-\frac{m_{i-1,j,k}}{m_{i,j,k}}\right)e^{\left(1-\frac{m_{i-1,j,k}}{m_{i,j,k}}\right)u_b}e^{\frac{m_{i-1,j,k}}{m_{i,j,k}}u_{i,j,k}}\right]\nonumber\\
&&-\frac{\tilde{c}^2}{2h^2}\left(6-\frac{m_{i-1,j,k}}{m_{i,j,k}}\right)b_{i,j,k}-\frac{\tilde{c}^2}{2h^2}\left(1-\frac{m_{i-1,j,k}}{m_{i,j,k}}\right)\frac{m_{i-1,j,k}}{m_{i,j,k}}e^{\left(1-\frac{m_{i-1,j,k}}{m_{i,j,k}}\right)u_b}e^{\frac{m_{i-1,j,k}}{m_{i,j,k}}u_{i,j,k}}\left(u_{i,j,k}-u_b\right)\nonumber\\
&&-\frac{1}{3(n+1)}\Omega_ma^4\left(na^2\xi\right)^{\frac{1}{n+1}}\exp\left(-\frac{u_{i,j,k}}{n+1}\right).
\end{eqnarray}
\end{widetext}
Note that when $m_{i-1,j,k}=0$ and $u_b=u_{i-1,j,k}$, the above equations reduce to those for a regular grid with periodic boundary conditions as expected. The relaxation on this level is then done through
\begin{eqnarray}
u^{h,{\rm new}}_{i,j,k} &=& u^{h,{\rm old}}_{i,j,k} - \frac{\mathcal{L}^h\left(u^{h,{\rm old}}_{i,j,k}\right)}{\frac{\partial\mathcal{L}^h(u^{h,{\rm old}}_{i,j,k})}{\partial u^{h,{\rm old}}_{i,j,k}}}.
\end{eqnarray}

\subsubsection{The Multigrid Implementation}

Let us write the BC-modified $f(R)$ equation as
\begin{eqnarray}
\mathcal{L}^h\left(u^h\right) &=& f^h
\end{eqnarray}
on the fine level, where $\mathcal{L}$ is the nonlinear operator acting on $u^h$ defined above. Note that here $f^h=0$, but we shall still keep it for a reason which will become clear below.

After a number of pre-smoothing iterations, one finds an approximate solution $\hat{u}^h$ on the fine level, which gives
\begin{eqnarray}
\mathcal{L}^h\left(\hat{u}^h\right)\ =\ \hat{f}^h\ \neq\ f^h.
\end{eqnarray}
Consider the new equation
\begin{eqnarray}
\mathcal{L}^h\left(u^h\right) - \mathcal{L}^h\left(\hat{u}^h\right)\ =\ f^h-\hat{f}^h\ \equiv\ -d^h.
\end{eqnarray}
After coarsifying and rearranging, we obtain the coarser-level equation as
\begin{eqnarray}
\mathcal{L}^H\left(u^H\right) &=& \mathcal{L}^H\left(\mathcal{R}\hat{u}^h\right)-\mathcal{R}d^h.
\end{eqnarray}
After a number of relaxation iterations on the coarser level, we find an approximate solution $\hat{u}^H$, and the fine-level solution can be corrected as
\begin{eqnarray}
\hat{u}^{h,{\rm new}} &=& \hat{u}^{h,{\rm old}} + \mathcal{P}\left(\hat{u}^H-\mathcal{R}\hat{u}^h\right),
\end{eqnarray}
where $\mathcal{P}$ is the prolongation operator.

Different from the linear case, here on the coarser level we are not solving an equation for the correction to the field $u$ (remember in that case we solved $\delta\phi$ on the coarser level) but again $u$ itself. Therefore Eqs.~(\ref{eq:bc_modified_fR_op}, \ref{eq:bc_modified_fR_dop}) could be applied to the coarse level as well, if we
\begin{enumerate}
\item change $h$ to $H$, and
\item find correct coarser-level boundary values for $u$, $u_B$.
\end{enumerate}
The first point is fairly straightforward, while the second needs some further analysis. One must first find the physical boundary on the coarser level which, as explained above, is where the coarser-level mask $m^H$ crosses zero. This is easy because $m^H$ is computed by restricting the fine-cell mask values. The values of $u_B$ in the boundary coarse cells are taken as those in the corresponding AMR cells.

Finally some comments should be made on the restriction operations $\mathcal{R}d^h$ and $\mathcal{R}\hat{u}^h$. $\mathcal{R}d^h$ will be treated similarly as in the linear case for the same reasons explained there.

For $\mathcal{R}\hat{u}^h$, even though a given fine cell is masked it still has contribution, because otherwise the computed $\mathcal{R}\hat{u}^h$ for the coarser cell will be incorrect. But if a coarser cell is masked we will not compute $\mathcal{R}\hat{u}^h$ for it, since this will be unused anyway, as in the case of $\mathcal{R}d^h$.

The truncation error $\tau^h$ can be estimated as
\begin{eqnarray}
\tau^h &\approx& \mathcal{L}^H\left(\mathcal{R}\hat{u}^h\right) - \mathcal{RL}^h\left(\hat{u}^h\right),
\end{eqnarray}
and similarly for other levels. This could provide a stopping (convergence) criterion for the multigrid iteration, which in our case is
\begin{eqnarray}
\left|d^h\right| \lesssim \alpha\left|\tau^h\right|,
\end{eqnarray}
where $\alpha\leq1/3$ is a predefined constant.

\subsubsection{The Prolongation}

As mentioned above, in the nonlinear case we have to prolongate the quantity $(\hat{u}^H-\mathcal{R}\hat{u}^h)$, but not the residual itself. The prolongated result can then be used to correct the fine-level solutions.

The prolongation here is done in the same way as that for the residual of the linear (modified) Poisson equation, to be consistent. Details shall not be presented here.

\section{Code Tests}

\label{sect:tests}

In this section we shall give the results of some tests we have performed to show that the above algorithm and implementation work correctly and efficiently.

\subsection{$f(R)$ Equation Solver On Domain Grid}

The most important part in the modified {\tt RAMSES} code, which is the topic of the last two sections, is the equation of motion for $f(R)$ gravity or the new degree of freedom(s) in other theories. Here we have performed a range of tests for it with different configurations of the matter density field.

\subsubsection{Homogeneous Matter Density Field}

In a universe with homogeneous density, we know that the quantity $f_R$ is homogeneous and it exactly takes its background value $\bar{f}_R$, namely
\begin{eqnarray}
\bar{\tilde{f}}_R &=& -\frac{na^2\xi}{\left[3\left(a^{-3}+4\frac{\Omega_\Lambda}{\Omega_m}\right)\right]^{n+1}},
\end{eqnarray}
everywhere. Therefore, as the simplest test of the $f(R)$ equation solver, one has to show that in such a homogeneous field, given some random guess of $\tilde{f}_R$ on the cells of the simulation mesh, after a reasonable number of Gauss-Seidel relaxation sweeps, the solution converges to the above background value. Such simple test have been used previously in \cite{bbdls2011,dlmw2011} to show that the {\tt MLAPM} solver for extra degrees of freedom works well.

\begin{figure}
\includegraphics[scale=0.36]{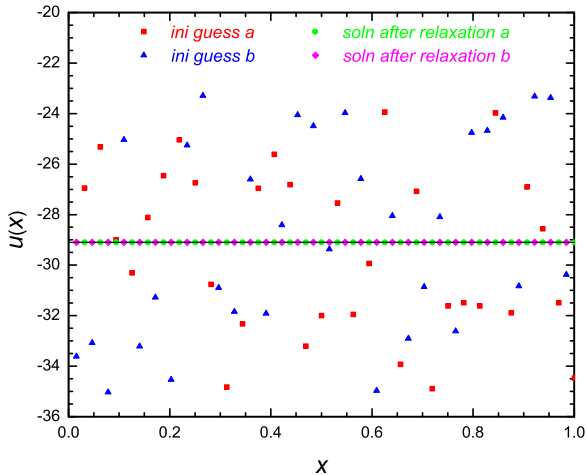}
\caption{(Colour online) Test of the solver for the $f(R)$ equation in a constant matter density field. Only results in the cells along the $x$-axis are shown, and the $x$-coordinate is rescaled by the size of the simulation box so that $x\in[0,1]$. Two initial guesses of $\tilde{u}\equiv\log(-\tilde{f}_R)$ have been tried (the red squares and blue triangles), the final answer corresponding to which are respectively denoted by green filled circles and purple diamonds. The black horizontal line is the exact solution. Note that the results with the first initial guess (filled squares and circles) have been shifted rightwards to make the plot clearer.} \label{fig:test_const_dens}
\end{figure}

We have performed this test for $a\approx0.04, n=1$ and a value of $\xi$ corresponding to $|f_{R0}|=10^{-5}$. The result is shown in Fig.~\ref{fig:test_const_dens}, where we plot the values of $\tilde{u}=\log(-\tilde{f}_R)$ in the cells in $x$-direction, before and after the Gauss-Seidel relaxation, for two different random initial guesses. We can see that the final solution agrees with the analytical result (the horizontal line) very well (see figure caption for more details).

\subsubsection{Point Mass}

As a second test of the {\tt RAMSES} $f(R)$ equation solver, let us consider the solution of $f_R$ around a point mass at the origin, for which case we have an analytical solution which is accurate except for the regions very close to the mass. Such a test has been used previously in \cite{o2008,bbdls2011}.

Following \cite{o2008}, we construct the point-mass density field as
\begin{equation}
\label{eq:point_mass}
\delta_{i,j,k} = \left\{%
\begin{array}{ll}
10^{-4}\left(N^3-1\right), & \hbox{$i=j=k=0$;} \\
-10^{-4}, & \hbox{otherwise.} \\
\end{array}%
\right.
\end{equation}
in which $i,j,k$ are respectively the cell indices in $x,y,z$ direction. In the test we have used a cubic box with size $256h^{-1}$Mpc and 128 grid cells in each direction. The other physical parameters are chosen as $a=1, n=1$ and we have used three values of $\xi$ corresponding to $|f_{R0}|=10^{-6}, 10^{-5}, 10^{-4}$.

Meanwhile, the analytical solution can be obtained approximately by solving the equation
\begin{eqnarray}
\nabla^2\delta f_R &\approx& m^2_{\rm eff}\delta f_R
\end{eqnarray}
in which the effective mass of the scalar degree of freedom $\delta f_R$, $m_{\rm eff}$, is given by
\begin{eqnarray}
m_{\rm eff} &=& \frac{1}{\sqrt{3}}\frac{1}{n(n+1)\xi}\left[3\left(1+4\frac{\Omega_{\Lambda}}{\Omega_m}\right)\right]^{n+2}\Omega_mH^2_0.
\end{eqnarray}

\begin{figure}
\includegraphics[scale=0.36]{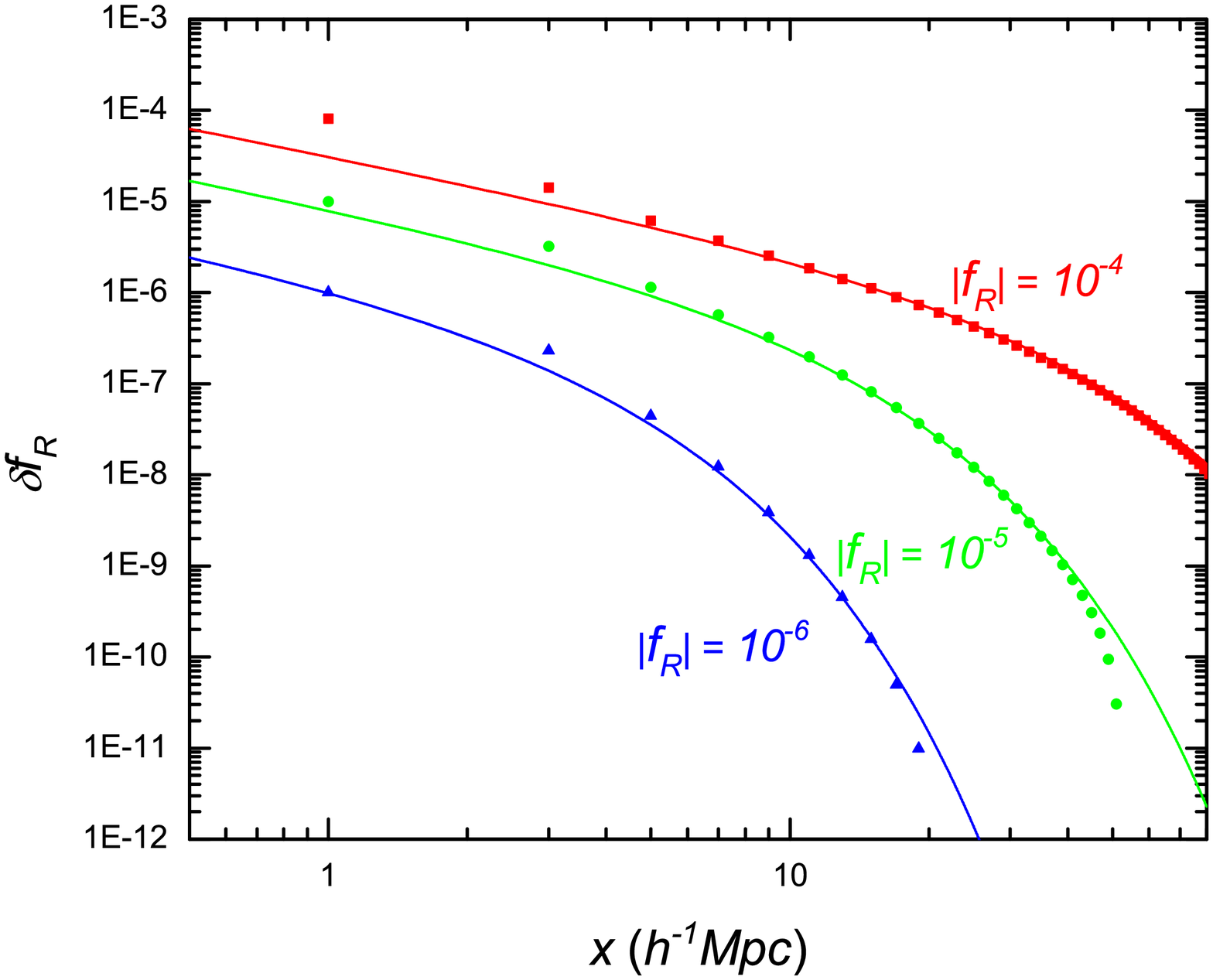}
\caption{(Colour online) The  solution to $\delta f_R\equiv f_R-\bar{f}_R$ around a pointed mass constructed according to Eq.~(\ref{eq:point_mass}), for three models with $|f_{R0}|=10^{-6}$ (blue triangles), $10^{-5}$ (green filled circles) and $10^{-4}$ (red squares) respectively. The solid curves are the corresponding analytical approximations which are accurately far from the point mass. Only the solutions along the $x$-axis are shown.} \label{fig:test_point_mass}
\end{figure}

Fig.~\ref{fig:test_point_mass} shows the comparison between the numerical solutions to $\delta f_R$ along the $x$-axis (symbols) and analytical solutions (solid curves), and we can see that the two agree very well for all three models used in this test.

\subsubsection{Sine Density Field}

As our third test, let us consider the sine density field introduced in \cite{o2008}, which (after some modification to account for the code units) is given by
\begin{eqnarray}\label{eq:sine_dens}
\delta(x) &=& \frac{\tilde{c}^2}{\Omega_ma}(2\pi)^2\frac{na^2\xi}{\left[3\left(a^{-3}+4\frac{\Omega_\Lambda}{\Omega_m}\right)\right]^{n+1}}\sin(2\pi x)\nonumber\\
&&+\left(1+4a^3\frac{\Omega_\Lambda}{\Omega_m}\right)\left\{\left[2-\sin(2\pi x)\right]^{-\frac{1}{n+1}}-1\right\},
\end{eqnarray}
in which $x$ is rescaled such that $x\in[0,1]$. Notice that we consider only $x$-dependence, which is equivalent to a one-dimensional configuration. The solution to this density field can be analytically worked out to be,
\begin{eqnarray}
\tilde{f}_R(x) &=& \frac{na^2\xi}{\left[3\left(a^{-3}+4\frac{\Omega_\Lambda}{\Omega_m}\right)\right]^{n+1}}\left[\sin(2\pi x)-2\right].
\end{eqnarray}

\begin{figure}
\includegraphics[scale=0.36]{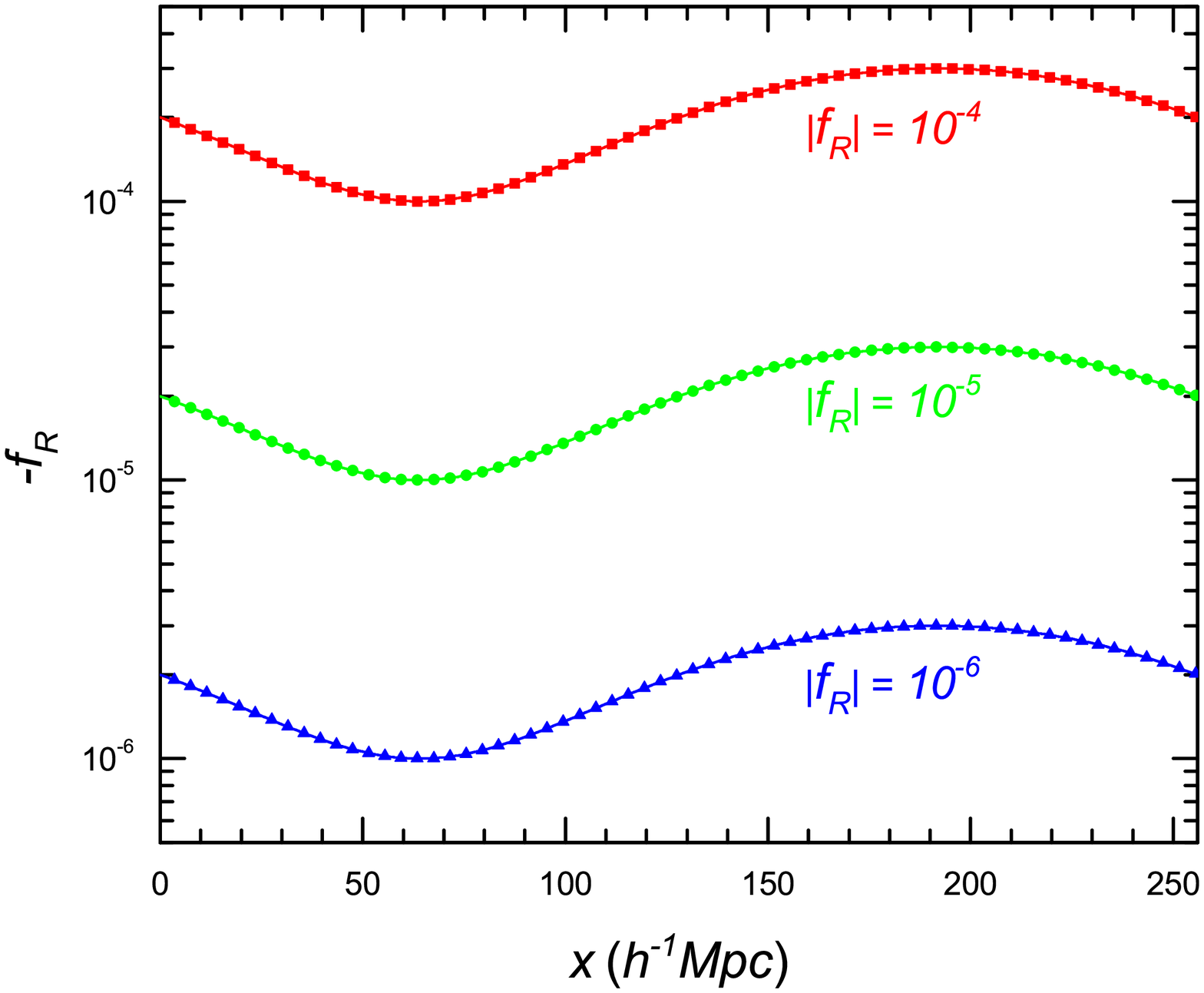}
\caption{(Colour online) Solutions of $f_R$ in a one-dimensional ($x$-direction) sine density field constructed using Eq.~(\ref{eq:sine_dens}), for three models with $|f_{R0}|=10^{-6}$ (blue triangles), $10^{-5}$ (green filled circles) and $10^{-4}$ (red squares) respectively. The solid curves are the corresponding analytical results. A simulation box with side length of $256h^{-1}$Mpc and 256 grid cells on each side is used in the computation. $x$ is in physical units.} \label{fig:test_sine_dens}
\end{figure}

Fig.~\ref{fig:test_sine_dens} shows the test results for the sine density field given above, with $a=n=1$ and three values of $\xi$ corresponding to $|f_{R0}|=10^{-6}, 10^{-5}, 10^{-4}$ respectively. It can be seen that the numerical solutions (symbols) agree with the analytical solutions (solid curves) very well. 

\subsubsection{Gaussian Density Field}

The last test of the $f(R)$ equation solver on the domain grid uses a Gaussian type density configuration. Again, here we only consider one-dimensional case, where the density field is specified by
\begin{eqnarray}\label{eq:gaussian_dens}
\delta(x) &=& \frac{c^2}{\Omega_ma}\frac{2A\alpha}{W^2}\exp\left[-\frac{(x-0.5)^2}{W^2}\right]\left[1-2\frac{(x-0.5)^2}{W^2}\right] \nonumber\\
&&+\frac{1}{3}a^3\left\{\frac{na^2\xi}{A\left[1-\alpha\exp\left(-\frac{(x-0.5)^2}{W^2}\right)\right]}\right\}^{\frac{1}{n+1}} \nonumber\\
&&-\left[1+4\frac{\Omega_\Lambda}{\Omega_m}a^3\right]
\end{eqnarray}
where again $x$ has been scaled to code units so that $x\in[0,1]$, $W$, $\alpha$ are numerical constants which respectively specify the width and height of the density field, which obviously peaks at $x=0.5$, and $A$ is defined by
\begin{eqnarray}
A &\equiv& \frac{na^2\xi}{\left[3\left(a^{-3}+4\frac{\Omega_\Lambda}{\Omega_m}\right)\right]^{\frac{1}{n+1}}}
\end{eqnarray}
Note that such a density field is not exactly periodic at the edges of the simulation box, but given that $W$ is small enough, $\delta\rightarrow0$ at the box edges and periodic boundary conditions are approximately satisfied.

\begin{figure}
\includegraphics[scale=0.36]{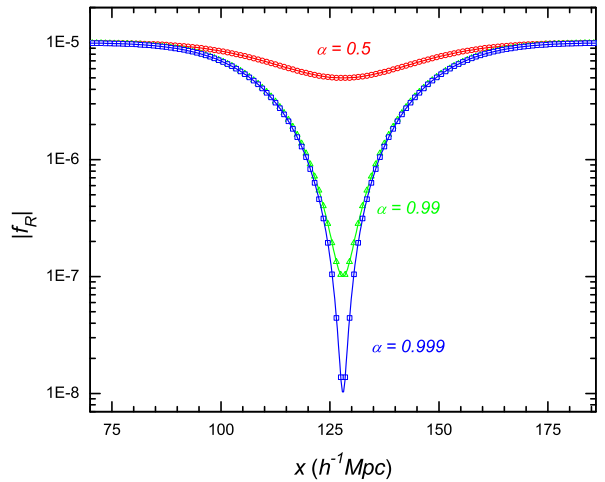}
\caption{(Colour online) Solutions of $f_R$ in a one-dimensional ($x$-direction) Gaussian-type density configuration constructed using Eq.~(\ref{eq:gaussian_dens}), for the models with $|f_{R0}|=10^{-5}$ and $\alpha=0.5$ (red circles), $0.99$ (green triangles) and $0.999$ (blue squares). The solid curves are the corresponding analytical results from Eq.~(\ref{eq:fR_gaussian_dens}). A simulation box with side length of $256h^{-1}$Mpc and 256 grid cells on each side is used in the computation and the $f(R)$ equation is only solved on the regular domain grid. $x$ is in physical units.} \label{fig:test_gaussian_dens}
\end{figure}

The solution to $f_R$ can then be obtained analytically and is
\begin{eqnarray}\label{eq:fR_gaussian_dens}
f_R &=& -A\left[1-\alpha\exp\left(-\frac{(x-0.5)^2}{W^2}\right)\right],
\end{eqnarray}
which clearly shows that when $\alpha\rightarrow1$ $|f_R|$ could be made very small at $x=0.5$ while at $x\rightarrow0$ or $x\rightarrow1$ $f_R$ goes to its background value.

We have implemented Eq.~(\ref{eq:gaussian_dens}) into our numerical code and the numerical solutions for $f_R$ are shown in Fig.~\ref{fig:test_gaussian_dens}. For simplicity we only show the results for $|f_{R0}|=10^{-5}$, but for three different values of $\alpha=0.5,0.99$ and $0.999$ (the open circles, open triangles and open squares in Fig.~\ref{fig:test_gaussian_dens} respectively). We can see that the numerical results agree with the analytical solution Eq.~(\ref{eq:fR_gaussian_dens}) very well.

\subsection{$f(R)$ Equation Solver On Refinements}

\begin{figure}
\includegraphics[scale=0.36]{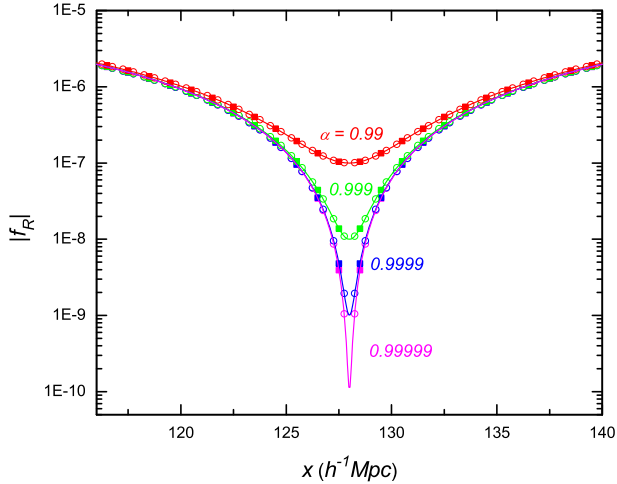}
\caption{(Colour online) The same as Fig.~\ref{fig:test_gaussian_dens}, but for the models with $|f_{R0}|=10^{-5}$ and $\alpha=0.99, 0.999, 0.9999, 0.99999$ (from top to bottom: red, green, blue, magenta). The $f(R)$ equation is solved on two levels: level 8 (the regular domain grid) and level 9 (the first refinement), and their numerical solutions are represented by filled squares and empty squares respectively. The solid curves are the corresponding analytical results from Eq.~(\ref{eq:fR_gaussian_dens}).} \label{fig:test_gaussian_dens2}
\end{figure}

The above four tests show that our solver for the $f(R)$ equation actually works accurately on the regular domain grid, but this is not sufficient for the code test since the $f(R)$ equation is also solved on refinements where, as we have seen, the equations take different forms due the complex treatment of the boundary conditions. It is therefore essential to test the $f(R)$ equation solver on the refinements as well, as we shall do in this subsection.

\subsubsection{Two-level Gaussian Density Field}

The Gaussian-type density configuration could provide a good check of the multilevel $f(R)$ equation solver because the density peak can be made arbitrarily high by adjusting the parameter $\alpha$. Near its peak, the density field changes rapidly and a higher spatial resolution is needed to compute $f_R$ accurately. Consider the case where the regular domain grid is refined only once, in regions where the density value exceeds some certain threshold (we shall call this a two-level problem, and in the present numeric example the coarse and fine levels are respectively levels 8 and 9). The density values in both the coarse and refined cells are given by Eq.~(\ref{eq:gaussian_dens}), while the values of $f_R$ at the fine-level boundaries are computed from interpolation of those in the nearby coarse-level cells, as discussed above.

Fig.~\ref{fig:test_gaussian_dens2} shows the numerical values of $f_R$ on both levels in the region covered by the refinement. We show the results for four different values of $\alpha$ ($0.99, 0.999$, $0.9999$, $0.99999$ from top to bottom), and for each $\alpha$ the results from the coarse and fine levels are denoted respectively by filled squares and empty circles. For comparison we have also plotted the analytical results Eq.~(\ref{eq:fR_gaussian_dens}) as solid curves. As we can see, both fine-level and coarse-level results are virtually indistinguishable from the exact solution by eye.

This does not mean that the refinement is unnecessary however, because, as shown in Fig.~\ref{fig:test_gaussian_dens2}, the fine level has more data points and could probe regions closer to the extreme value of $f_R$, which corresponds to the high-density region where high resolution is needed.

\begin{figure}
\includegraphics[scale=0.33]{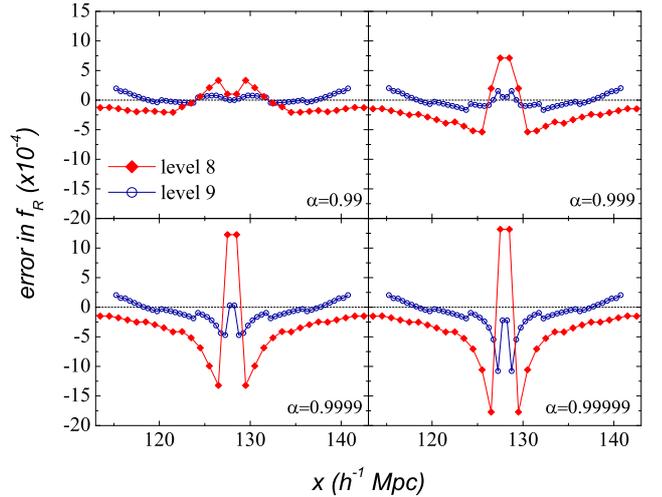}
\caption{(Colour online) Numerical errors of $f_R$ (defined as the fractional difference between the numerical result and the exact solution Eq.~(\ref{eq:fR_gaussian_dens})) for the models with $|f_{R0}|=10^{-5}$ and $\alpha=0.99$ (upper left panel), $0.999$ (upper right), $0.9999$ (lower left) and $0.99999$ (lower right). The $f(R)$ equation is solved on two levels: level 8 (the regular domain grid, red filled diamonds) and level 9 (the first refinement, blue hollow circles).} \label{fig:test_gaussian_dens3}
\end{figure}

To see more quantitatively how well the solutions from different levels agree with the exact results, we have plotted in Fig.~\ref{fig:test_gaussian_dens3} the percentage errors of the numerical solutions for the same four models considered in Fig.~\ref{fig:test_gaussian_dens2}. The following features could be easily observed from this plot:
\begin{enumerate}
\item At the coarse level (level 8), the numerical error increases as the density fluctuation becomes stronger (i.e., $\alpha\rightarrow1$), which is as expected;
\item Around the peak of the density field, the fine level (level 9) consistently gives more accurate numerical results than level 8 for all 4 models, thanks to its higher spatial resolution;
\item The improvement of level-9 result over that of level-8 is smaller as $\alpha\rightarrow1$, because at the peak of the density field even level 9 will not be very accurate;
\item The level-9 result is actually less accurate at the refinement boundaries, because of the numerical error that comes from setting boundary conditions using interpolation. Furthermore, the error at refinement boundaries is the same for all models, at $\sim0.025\%$, which is negligible anyway. Note that we could improve on this using a higher-order interpolation scheme at boundaries, which will be considered in future work.
\end{enumerate}

\subsubsection{Three-level Gaussian Density Field}

\begin{figure}
\includegraphics[scale=0.22]{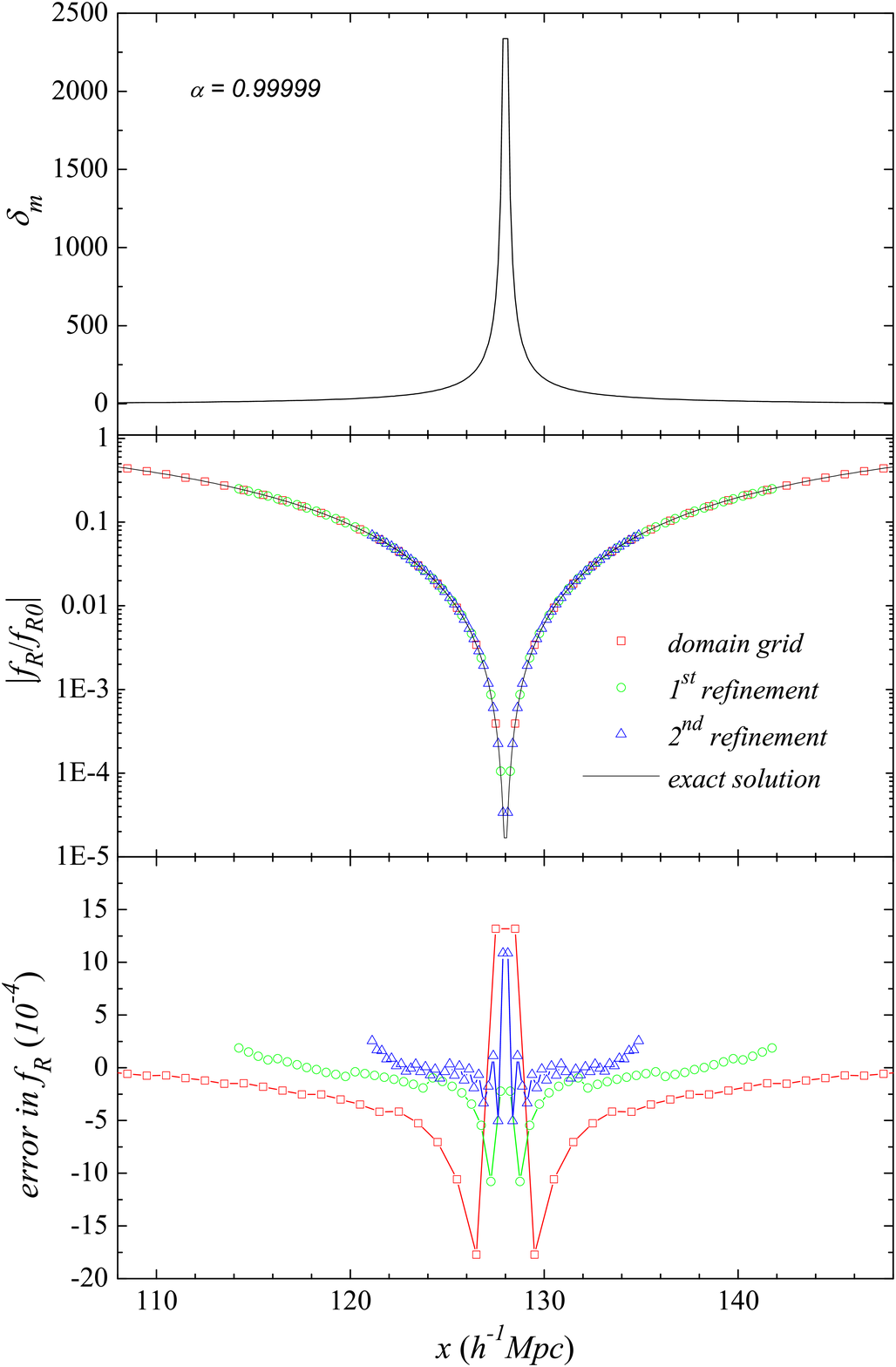}
\caption{(Colour online) {\it Upper Panel:} The one-dimensional matter density field as a function of the $x$ coordinate (in physical units), to provide an impression about the height of its peak. $\delta_m\equiv\rho_m/\bar{\rho}_m-1$.  {\it Middle Panel:} The same as Fig.~\ref{fig:test_gaussian_dens2}, but for the model with $|f_{R0}|=10^{-5}$ and $\alpha=0.99999$ only for simplicity. The $f(R)$ equation is solved on three levels: level 8 (the regular domain grid) and level 9 (the first refinement) and level 10 (the second refinement), and their numerical solutions are represented by red squares, green circles and blue triangles respectively. The black solid curve is the corresponding analytical result from Eq.~(\ref{eq:fR_gaussian_dens}). {\it Lower Panel:} The same as Fig.~\ref{fig:test_gaussian_dens3}, but for the model studied here; the use of symbols is the same as in middle panel.} \label{fig:test_gaussian_dens4}
\end{figure}

In the above we have performed a test using the domain grid and one refinement level, but the generalisation to more refinement levels is very straightforward (and indeed done automatically in the code, with the refinement criterion specified {\it a priori}). Here we only give an example with three refinement levels, again making use of the above Gaussian-type density field.

Fig.~\ref{fig:test_gaussian_dens4} shows the results of the three-level test. In the upper panel we have plotted the density field $\delta_m$ as function of $x$, and it could be seen that it has a rather high and sharp peak near $x=128h^{-1}$Mpc, which is exactly where higher spatial resolution and thus refinements are needed. In the middle panel we have shown the numerical solutions from the three refinement levels (8, 9 and 10) in the spatial regions covered by them, and again we see that the agreements with the exact solution are very good (indistinguishable by eye). The numerical errors from the $f(R)$ equation solver at the three levels are shown in the lower panel, and we see the same patterns as observed in Fig.~\ref{fig:test_gaussian_dens3}, namely near the density peak higher refinement level produces smaller error, but at the refinement edges slightly bigger error occurs because of the interpolation used in setting the boundary conditions. The magnitude of the latter error source, however, is much smaller than that near the density peak, and is negligible for cosmological simulations anyway.

These tests indicate that our $f(R)$ equation solver is reasonably accurate and therefore reliable also on the refinement(s), which is one of the most significant achievements of the present code.

\begin{table*}
\label{tab:codes_summary} \caption{A brief summary of the key features of the three known grid-based codes for $N$-body simulations in modified gravity and/or dynamical dark energy theories. Here {\tt OC} stands for {\it Oyaizu Code}, and by {\tt MLAPM} we mean the modified version to include the multigrid solver for the extra degree(s) of freedom.}
\begin{tabular}{@{}lccc}
\hline\hline
Codes & {\tt OC} & {\tt MLAPM} & {\tt ECOSMOG}\\
\hline
Reference & Ref.~\cite{o2008} & Refs.~\cite{lz2009,lb2011} & This paper\\
Density assignment scheme & CIC & TSC & TSC\\
Force interpolation scheme & CIC & TSC & TSC\\
Parallelisation & {\tt OpenMP} & N/A & {\tt MPI}\\
Adaptive refinement? & No & Yes & Yes \\
Multigrid on regular grid? & Yes & Yes & Yes\\
Multigrid relaxation arrangement on regular grid\ \ \ \ \ \ & V-cycle\ \ \ \   & V-cycle/adaptive cycle\ \ \ \  & V-cycle\\
Multigrid on refinement? & N/A & No & Yes\\
Multigrid relaxation arrangement on refinement\ \ \ \ \ & N/A & N/A & V-cycle\\
Programming language & {\tt C++} & {\tt C} & {\tt FORTRAN 90} \\
\hline\hline
\end{tabular}
\end{table*}

\subsection{Density Assignment and Force Interpolation}

\begin{figure}
\includegraphics[scale=0.32]{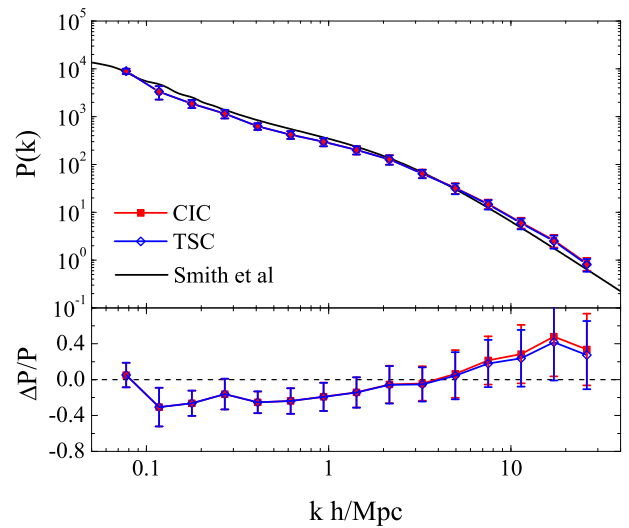}
\caption{(Colour online) {\it Upper Panel:} Comparison of the nonlinear matter power spectra using the TSC (blue diamonds) and CIC (red squares) density assignment schemes to that obtained using Smith {\it et~al}.~fit (black solid curve). {\it Lower Panel:} The relative differences of the TSC and CIC power spectra with respect to that of the Smith {\it et~al.}~fit. The dashed black line is zero identically. The cosmological simulations here use a cubic box with size $100h^{-1}$Mpc and $256$ cells on each side of the domain grid. The error bars show the sample variance over 5 different realisations. All results are at redshift 0.} \label{fig:test_tsc_cic}
\end{figure}

The default {\tt RAMSES} code uses the CIC (cloud-in-cell) scheme to do density assignment and force interpolation. While the CIC scheme works well and is the most widely used due to a compromise of accuracy and cost, the TSC (triangle-shaped clouds) scheme actually produces a smoother density field which more resembles the true dark matter distribution, especially when the particle number is not large enough. Furthermore, this results in a more isotropic force around a point mass, a desirable property. For this reason, we have added a new routine in the code to do the TSC density assignment. Correspondingly, the force interpolation also needs to use the TSC scheme to ensure momentum conservation. For more discussions on the different density assignment schemes we refer the readers to \cite{hebook}.

Because in the TSC density assignment scheme a particle's cloud \footnote{In density assignment, a particle only contributes to the density field within a finite region around it, and this is called that particle's cloud. The particle's contribution to density field integrated over its cloud is its mass. As a result, the bigger the cloud is, the smoother the resulted density field will be.} spreads more than in the CIC scheme, the resulted density field will be smoother and as a result we expect less small-scale structure from $N$-body simulations. This is confirmed by our numerical results shown in Fig.~\ref{fig:test_tsc_cic}, where we have plotted the nonlinear matter power spectra from our $\Lambda$CDM runs using TSC and CIC respectively, and compared them with the Smith {\it et~al.} fit \cite{Smith:2002dz}. We can see that on very large scales (small $k$) TSC gives virtually the same prediction as CIC, but on smaller scales (large $k$) it produces less power than the latter. On the other hand, the difference between TSC and CIC (a few percent) is much smaller than their difference from the Smith {\it et~al.} fit (up to 40\% on small scales).

A simpler test of the code segments for the TSC density assignment is to check that the average density of all grid cells on the domain grid is $1.0$. We have monitored this at each time step of all our cosmological simulations and found that this is indeed the case. As a result, we believe that this part of our {\tt ECOSMOG} code is reliable.

\subsection{Performance Tests}

In addition to a more accurate $f(R)$ equation solver, another aim we want to achieve for our code is efficiency. The efficiency in the $f(R)$ equation solver can be greatly improved by two factors. The first (as mentioned earlier) is to use multigrid rather than a single grid in arranging the Gauss-Seidel relaxation to speed up its convergence, and the other is using parallelisation to take advantage of the supercomputing resources. These two elements are incorporated in the original {\tt RAMSES} code for the Poisson solver, and in our code we apply them to the $f(R)$ equation solver as well.

\subsubsection{Performance of V-cycle}

\begin{figure}
\includegraphics[scale=0.36]{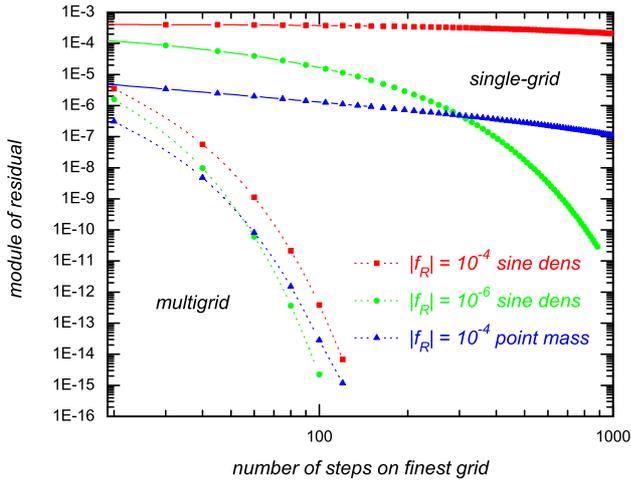}
\caption{(Colour online) The reduction of the residual for the $f(R)$ equation for three models: $|f_{R0}|=10^{-4}$ with a sine-type density field as discussed above (red squares), $|f_{R0}|=10^{-6}$ with a sine-type density field (green circles) and $|f_{R0}|=10^{-4}$ for a point mass discussed above (blue triangles). The results using a single grid are represented by symbols connected by solid curves, while those using the multigrid method (arranged by V-cycles) by symbols connected by dashed curves.} \label{fig:test_mg_performance}
\end{figure}

We use a V-cycle to arrange the multigrid relaxation, as in the modified {\tt MLAPM} code \cite{lz2009,lb2011}. As a check of the improvement that is achieved in this way, we have plotted in Fig.~\ref{fig:test_mg_performance} the convergence rates for three models (details of which can be found in the figure caption) using single-grid relaxation (solid curves) and multigrid V-cycle (dashed curves). The convergence rate is equivalent to the rate at which the module of the residual decreases. We can see that with the multigrid relaxation the residual drops below $10^{-12}$ after tens of Gauss-Seidel sweeps on the finest grid, while with the single-grid relaxation this depends sensitively on both the model and its parameters -- in the worst case we have tested here (the sine density field with $f_{R0}=-10^{-4}$, solid curve with filled squares in Fig.~\ref{fig:test_mg_performance}) the residual is still bigger than $10^{-4}$ even after 1000 Gauss-Seidel sweeps!\footnote{Of course this also depends on the initial guess -- if it is close enough to the true solution then rapid convergence could be expected. But in most situations we have no idea about the exact solution and cannot find such good initial guesses.} Clearly the single-grid relaxation is impractical in most realistic simulations.

\begin{figure}
\includegraphics[scale=0.36]{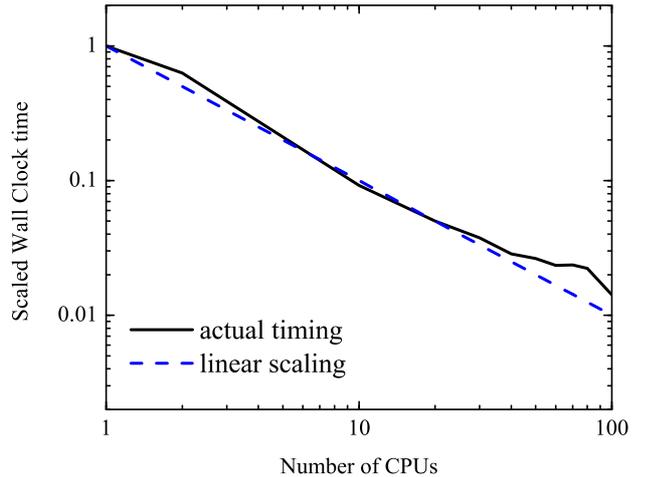}
\caption{(Colour online) The wallclock time of the $f(R)$ equation solver as a function of the number of CPUs used in the parallelised computation. The actual measurement is shown as the black solid curve, while the blue dashed line is a linear scaling just for comparison. It is obvious that the time-CPU number scaling is approximately linear for CPU numbers up to $\sim100$.} \label{fig:test_wtime_ncpu}
\end{figure}

One difference between the code here and the modified {\tt MLAPM} \cite{lz2009, lb2011} is that here the multigrid relaxation method is used not only on the regular domain grid, but also on the refinements. In contrast, in {\tt MLAPM} single-grid relaxation is used on refinements, which gains ease in the code implementation but at the cost of computing time and accuracy. Tests of our code show that on the refinements the V-cycle could bring the residual down to $10^{-12}$ within a reasonable number of Gauss-Seidel sweeps, just as it does on the regular grid. Therefore, we expect more accurate results than those obtained in previous works; we will come back to this in a future work.

\subsubsection{Parallelisation Performance}

Parallelisation is a way to let many CPUs share the computational overhead and therefore reduce the total physical time needed to finish the job. It has greatly improved the efficiency of numerical simulations in cosmology. Our parallelisation here uses the structure of the original {\tt RAMSES} code, which is parallelised using {\tt MPI} and has been shown to work very well for the Poisson equation solver. This is made possible by the fact that the mesh structures used for the Poisson and $f(R)$ equations are exactly the same, and so the structures for communication (which is a crucial ingredient of parallelisation) could be shared by the Poisson and $f(R)$ equation solvers, provided that they will not be modified until both equations are solved and relevant quantities appropriately stored.

To check the efficiency of the parallelisation, we have used different numbers of CPUs to solve the $f(R)$ equation (only once) and recorded the wallclock times used. This is plotted in Fig.~\ref{fig:test_wtime_ncpu} from which we can see that, at least when the CPU number ranges between $1$ and $\sim100$, the wallclock time roughly scales linearly with the number of CPUs. This shows that the parallelisation does help to make the code faster, and therefore most suitable for future simulations with large numbers of particles, large box sizes and high spatial resolutions.

\subsection{Cosmological Simulations}

\begin{figure}
\includegraphics[scale=0.36]{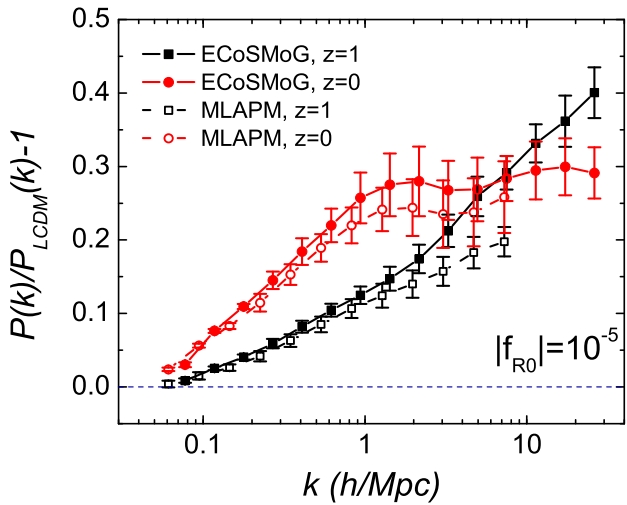}
\caption{(Colour online) Nonlinear power spectra at redshifts 1 (black filled squares) and 0 (red filled circles), from cosmological simulations for the $f(R)$ model with $|f_{R0}|=10^{-5}$. The horizontal axis is the wave number in the unit of $h$Mpc$^{-1}$, and the vertical axis is the fractional difference between the $f(R)$ and $\Lambda$CDM results. We have considered five realisations for each model, using the same set of initial conditions, box size ($100h^{-1}$Mpc), spatial resolution ($256$ cells on each side for the domain grid) and refinement criterion. The power spectra are averaged and the standard deviations are shown as error bars. We have also binned the horizontal axis to smooth the result \cite{zlk2011}. The corresponding results from \cite{zlk2011} have been plotted as black empty squares and red empty circles for comparison.} \label{fig:test_Pk_F5_B100}
\end{figure}

The aim of the modified code is to run cosmological simulations for modified gravity and dynamical dark energy theories, $f(R)$ gravity as an example. Therefore, we still need to test its reliability in such simulations, which is the purpose of this subsection.

As one cosmological test, we have performed simulations for the $f(R)$ gravity model with $|f_{R0}|=10^{-5}$ using a $100h^{-1}$Mpc simulation box and measured their matter power spectra. In order to reduce the cosmic variance, we show the matter spectrum for $f(R)$ gravity rescaled by that for a $\Lambda$CDM model simulated using the same initial condition. To reduce the sample variance, we make bins uniform in logarithm of the wavenumber $k$.

\begin{figure}
\includegraphics[scale=0.36]{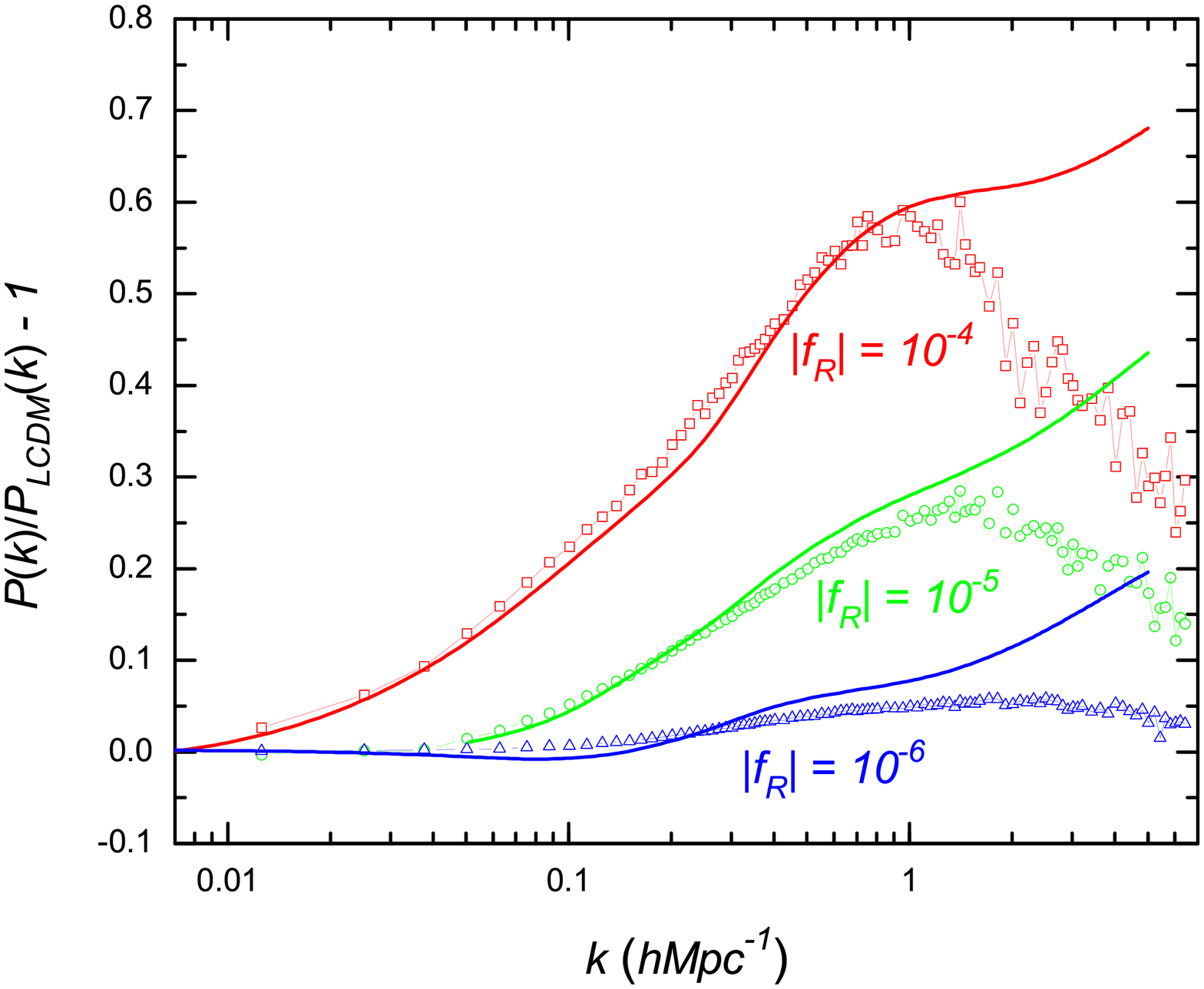}
\caption{(Colour online) Nonlinear matter power spectra at redshift 0, from cosmological simulations for the $f(R)$ models with $|f_{R0}|=10^{-4}$ (red empty squares), $10^{-5}$ (green empty circles) and $10^{-6}$ (blue empty triangles). The horizontal axis is the wave number in the unit of $h$Mpc$^{-1}$, and the vertical axis is the fractional difference between the $f(R)$ and $\Lambda$CDM results. The solid curves are the corresponding results from Smith {\it et~al.} fit.} \label{fig:test_Pk_F456_B500}
\end{figure}

The numerical results are summarised in Fig.~\ref{fig:test_Pk_F5_B100} (more technical details are described in the caption). To make comparison with the $\Lambda$CDM result, displayed in Fig.~\ref{fig:test_Pk_F5_B100} are the relative differences between the $f(R)$ and $\Lambda$CDM power spectra. At redshift $z=1$ ($a=0.5$), the fractional difference monotonically increases towards smaller scales, which have already been within the reach of the fifth force and started experiencing the boosted matter clustering; at redshift $0$ ($a=1.0$) the fractional difference goes down at smaller scales, because the particles have been accelerated, which helps to erase the small-scale structure to a certain degree \cite{zlk2011}. 

{We have also compared the results from {\tt ECOSMOG} and those from the modified {\tt MLAPM} code (empty symbols) in Fig.~\ref{fig:test_Pk_F5_B100}, which shows that the two codes agree quite well on large scales. On smaller scales, the quantitative agreement becomes worse but the qualitative features are still the same. In particular, the results at $z=0.0$ are within the $1\sigma$ error bars of each other. The discrepancy on small scales comes from the fact that the {\tt ECOSMOG} code could achieve much higher accuracy than {\tt MLAPM}: on the domain grid (with 256 cells in our {\tt ECOSMOG} runs and 128 cells in the {\tt MLAPM} runs) the convergence criterion is  $|d^h|<10^{-12}$ for the former code and $|d^h|<\max\left(10^{-6}, 0.03|\tau^h|\right)$ for the latter; on the refinements it becomes $|d^h|<10^{-8}$ and $|d^h|<\max\left(10^{-6}, 0.03|\tau^h|\right)$ for the two codes\footnote{We note that on refinements it is almost always the case that $10^{-6}\ll0.03|\tau^h|$, because the truncation error can be quite big on the fine levels. This means that the {\tt MLAPM} $f(R)$ equation solver was deemed as having converged even when $|d^h|\sim\mathcal{O}(0.01-1)$ on the higher refinements, while the {\tt ECOSMOG} solver has a much more stringent criterion of $|d^h|<10^{-8}$, and naturally will give more accurate results. Without using multigird relaxation on the refinements, such an accuracy will not become practical.}.}

{Because $|f_R|\ll1$ appears in the denominator of the right-hand side of the Poisson equation, a small difference in $f_R$ can be magnified and result in a big difference in the solution of the Newtonian potential. This is why a high accuracy in the solution of $f_R$ is important, and can also explain why the difference between the {\tt ECOSMOG} and {\tt MLAPM} results in Fig.~\ref{fig:test_Pk_F5_B100} is bigger at earlier times (when $|f_R|$ is smaller and its accuracy more important). Indeed, we find the small-scale structure in the $f(R)$ gravity to be quite sensitive to the convergence criterion of the $f(R)$ equation solver as well as the size of the simulation box, the refinement criterion {\it etc.}, and a thorough study needs to be conducted to clarify these issues before systematic simulations of the $f(R)$ gravity are performed. We leave this as a future work.} 

As the other cosmological test, we have also run simulations for three $f(R)$ gravity models with $|f_{R0}|=10^{-6}$, $10^{-5}$ and $10^{-4}$ respectively, for a bigger simulation box ($500h^{-1}$Mpc now). Because the measured matter power spectra are much smoother than the case above, here we shall not average over realisations or perform the binning. Fig.~\ref{fig:test_Pk_F456_B500} shows the enhancements of the power spectra in these models (symbols), and also the results using Smith {\it et~al.} fit (dashed curves). We could see that on the large scales they agree reasonably (for $k\ll1h$Mpc$^{-1}$ the difference is at most a few percent). This serves as another check of the consistency and reliability of our code.

As a final note, we find that simulations using bigger boxes generally takes less time to complete, provided that the particle numbers are the same. This is because the clustering is less developed and the cell size is larger, so that smaller scales are not resolved as well as for smaller box sizes. As a consequence, time steps are larger, and there are few refinement levels, even with the same refinement criterion. The density field is smoother, which makes the $f(R)$ equation solver converge more quickly. With 80 CPUs, our simulations with $500h^{-1}$Mpc box size can finish within a couple of hours, in contrast to about 24 hours for the simulations with $100h^{-1}$Mpc box size.

\section{Summary and Conclusions}

\label{sect:summary}

Large $N$-body simulations for modified gravity or dynamical dark energy theories have become more and more important in the wake of the inflow of high-quality observational data in the near future. To best extract information from these data, one needs better theoretical understandings of different models, in particular on scales of galaxy cluster and smaller sizes, where the commonly-used linear perturbation analyses are no longer valid.

In this work we have presented a new, efficiently parallelised, accurate and fast code called {\tt ECOSMOG} based on the public code {\tt RAMSES}. The {\tt ECOSMOG} code is specifically designed for performing large $N$-body and hydrodynamic simulations for modified gravity and generic dark energy theories with additional scalar degree(s) of freedom. We have clarified a number of numerical and technical issues about implementing the (usually nonlinear) equations of motion for those new degrees of freedom, and preformed a series of code tests, using $f(R)$ models, for the issues of the accuracy of the multigrid scalar field solver on the domain grid, on the refinements, the density assignment and the force interpolation scheme, as well as the efficiencies of the scalar field solver and of the parallelisation. Our results show that {\tt ECOSMOG} works extremely well. 

{\tt ECOSMOG} closely follows the convention and style of the original {\tt RAMSES} code. Special efforts have been made to not mess the default code, and the scalar field solver has been designed as several additional {\tt F90} files which are maximally parallel to the {\tt RAMSES} Poisson solver. There are some modifications to the default code, mainly to ensure smooth interfaces (such as make sure useful quantities are not destroyed between the calls to the scalar field and Poisson solvers), but these have been kept to minimal level.

{\tt ECOSMOG} is easy to use and modify for other gravity or dark energy models, making it a potentially powerful tool for massive $N$-body and hydrodynamical simulations for interesting theories of gravity and dynamical dark energy.

\begin{acknowledgments}
BL is supported by Queens' College and the Department of Applied Mathematics and Theoretical Physics of University of Cambridge. GBZ and KK are supported by STFC grant ST/H002774/1. KK also acknowledges support from the European Research Council and the Leverhulme Trust. BL thanks Drs.~David Mota and Hongsheng Zhao for encouragement to do this work, and the Institute of Cosmology and Gravitation for its host when this work was initialised. The test runs discussed in this paper have been performed on the {\tt SCIAMA} supercomputer of University of Portsmouth; the initial conditions for the cosmological runs were generated using the {\tt MPgrafic} code \cite{mpgrafic} and the matter power spectra were measured using {\tt POWMES} \cite{powmes}.
\end{acknowledgments}

\end{document}